\documentclass[11pt]{article}

\newcommand{\BE}{\begin{equation}}
\newcommand{\EE}{\end{equation}}
\newcommand{\be}{\begin{equation}}
\newcommand{\ee}{\end{equation}}
\newcommand{\BA}{\begin{eqnarray}}
\newcommand{\EA}{\end{eqnarray}}
\newcommand{\half}{{\scriptstyle{\frac{1}{2}}}}

\newcommand{\vol}{{\sf V}}
\newcommand{\num}{{\sf N}}

\newcommand{\delv}{\vec{\nabla}}
\newcommand{\delsq}{\nabla^2}
\newcommand{\dtr}{\delta^{(3)}({\bf r})} 
\newcommand{\mfp}{\lambda_{\rm mfp}}

\usepackage{amssymb}
\usepackage[dvips]{graphicx}

\begin{document}
\begin{titlepage}
\begin{flushright}
{\small DE-FG05-97ER41031-54 }
\end{flushright}
\vspace*{11mm}
\begin{center}
   {\LARGE{\bf Vacuum Hydrodynamics}}

\vspace*{15mm}

{\Large P. M. Stevenson}
\vspace*{3mm}\\
{\large T. W. Bonner Laboratory, Physics Department \\
Rice University, P.O. Box 1892, Houston, TX 77251-1892, USA}
\vspace{18mm}\\
{\bf Abstract:}
\end{center}

The Higgs vacuum -- with its constant background field -- is not `empty' 
but is a kind of medium.  
Any ordinary medium, viewed at sufficiently long length scales, has a  
hydrodynamic description and can propagate sound waves. 
The vacuum medium is unusual; the speed of sound is formally {\it infinite} 
and there is no linear sound-wave regime.  Instead, long-wavelength 
disturbances are described by some intrinsically nonlinear hydrodynamic 
equations.

\end{titlepage}

\newpage

     The purpose of this Letter is more to raise questions than to answer 
them.  The premise is that the vacuum of the Standard Model, with its 
non-zero background field $\langle \phi \rangle \neq 0$, is a kind of 
medium \cite{medium}.  The only other assumption is that a ``mean free path'' 
scale $\mfp$ exists and is finite.  (While $\mfp$ must be long compared to 
particle-physics length scales, whether it is a millimeter or an astronomical 
or cosmological scale is another question.)  My point is that at sufficiently 
long length scales $L \gg \mfp$ any fluid medium can be described by 
hydrodynamics (see, for example, \cite{landau,huang,degroot}).  That is, 
it can be treated as an ideal fluid up to corrections of order $\mfp/L$ 
\cite{fnteCE}.  

     In the hydrodynamic regime the only relevant modes are those  
directly linked to conservation laws.  In particular, energy-momentum 
conservation equations lead directly to longitudinal pressure waves --- 
sound waves.  Let us briefly review the derivation.  The energy-momentum 
tensor for a perfect fluid has the form [$c=1$, 
$g^{\mu\nu}= {\rm diag}(1,-1,-1,-1)$]
\BE
      T^{\mu \nu} = ({\cal E} + P)u^\mu u^\nu - P g^{\mu \nu},
\EE
where ${\cal E}$ is the energy density, $P$ is the pressure, and $u^\mu$ is 
the flow 4-velocity 
\BE
       u^\mu = (u_0,\vec{u}), \quad \quad \mbox{{\rm with}} \quad 
u_0^2-\vec{u}^2=1.
\EE
The energy-momentum conservation equation $\partial_\mu T^{\mu \nu}=0$ yields 
\BE
\label{em1}
\frac{\partial}{\partial t}\left[ ({\cal E} + P)u_0^2 \right] + 
\frac{\partial}{\partial x^j}\left[ ({\cal E} + P)u^j u_0 \right] - 
\frac{\partial P}{\partial t} = 0,
\EE
\BE
\label{em2}
\frac{\partial}{\partial t}\left[ ({\cal E} + P)u_0u^i \right] + 
\frac{\partial}{\partial x^j}\left[ ({\cal E} + P)u^j u^i \right] +  
\frac{\partial P}{\partial x^i} = 0.
\EE
Consider a small perturbation about a static, homogeneous equilibrium 
state ${\cal E}_0, P_0, \vec{u}=0$.  If the pressure variation $\delta P$ 
and the energy variation $\delta {\cal E}$ are sufficiently small they must 
be proportional, with their ratio fixed by the thermodynamic derivative 
\BE
\label{vs}
 \left. \frac{\partial P}{\partial {\cal E}}\right|_0 \equiv v_s^2.
\EE
Let $\psi=\psi({\bf x},t)$ denote the pressure variation $\delta P$.  
Substituting $P = P_0 + \psi$ and ${\cal E} = {\cal E}_0 + \psi/v_s^2$ into 
the equations above, keeping only terms first order in $\psi$ and/or 
$\mid \! \vec{u} \! \mid$, yields 
\BE
\label{enmom1}
   \frac{1}{v_s^2} \frac{\partial \psi}{\partial t} + 
   {\cal D} \delv.\vec{u}  =  0,
\EE
\BE
\label{enmom2}
   \delv \psi + {\cal D} \frac{\partial \vec{u}}{\partial t} 
   = 0,  
\EE
where ${\cal D} \equiv {\cal E}_0 + P_0$.  (For a nonrelativistic fluid 
${\cal D}$ would be the mass density.)  Taking $\partial/\partial t$ of 
the first equation minus $\delv.$ the second yields 
\BE
    \frac{1}{v_s^2} \frac{\partial^2 \psi}{\partial t^2} - \delsq \psi = 0,
\EE
which is the wave equation with wave velocity $v_s$. 

     Thus, the speed of sound is given by the thermodynamic derivative in 
Eq. (\ref{vs}).  For a classical gas, $v_s$ reflects the average speed of 
the molecules and vanishes at zero temperature.  However, a quantum gas of 
bosons has $v_s \neq 0$ even at $T=0$.  For zero temperature the energy 
density ${\cal E}$ is a function only of the number density $n$ and the 
pressure is given by 
\BE
P = -{\cal E} + n \frac{d {\cal E}}{dn},
\EE
so we may express $v_s^2$ as
\BE
\label{vs2}
v_s^2 =  n^2 \left. \left. \frac{d^2{\cal E}}{dn^2} \right|_0 \right/
 n \left. \frac{d{\cal E}}{dn} \right|_0 \equiv {\cal C}^{-1}/{\cal D}.
\EE
Here ${\cal C}$ is the compressibility (${\cal C}^{-1}$ is the adiabatic bulk 
modulus) and ${\cal D} = {\cal E}_0 + P_0$, as before.  For a nonrelativistic 
dilute Bose gas the ${\cal E}(n)$ function is  
\BE
\label{dbg}
{\cal E} = n m + n^2 \left( \frac{2 \pi a \hbar^2}{m} \right),
\EE
where $a$ is the scattering length.  The first term is just the rest-energy  
contribution ($c=1$); it dominates the energy density nonrelativistically, but 
cancels in the pressure.  The second term arises from the 2-body 
interactions.  From the last two equations one obtains the familiar 
Bogoliubov result $v_s^2 = 4 \pi n a \hbar^2/m^2$ \cite{huang}.  

      An `empty' vacuum could be viewed as the $n \to 0$ limit of the 
preceding case.  It is uninteresting for two reasons: the mean free path 
$1/(na^2)$ necessarily tends to infinity, and the speed of sound tends to 
zero.  

     However, a Higgs-type vacuum is quite different. The energy density is 
minimized, not at $n=0$, but at a local minimum, $n=n_v$.  (See Fig. 1.)  
Thus, $d{\cal E}/dn$ vanishes in the vacuum.  It follows directly that 
${\cal D}=0$ and $v_s^2$ is {\it infinite}!

\begin{figure}[htp]
\begin{center}
\includegraphics[width=6.0cm]{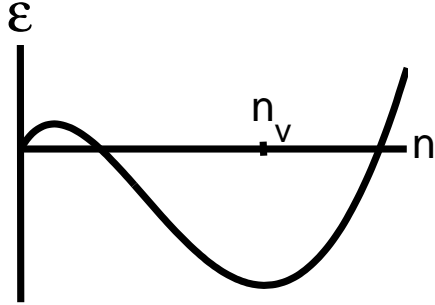}
\caption{The form of ${\cal E}(n)$ in the Higgs-vacuum 
case.  The vacuum is at $n=n_v$ where $d{\cal E}/dn = 0$. }
\end{center}
\end{figure}

     However, the linearized equations (\ref{enmom1}, \ref{enmom2}) 
are valid only when $\psi \ll {\cal D}$.  For any ordinary medium there is 
always such a linear regime, but for the vacuum ${\cal D}=0$.  Thus, we must 
return to the full equations.  The fact that $v_s^2=\infty$ in the Higgs 
vacuum really means that $\delta {\cal E}$ and $\delta P$ are no longer 
proportional to each other.  While $\delta P$ is linear in $\delta n$, the 
energy-density change $\delta {\cal E}$ is second order in $\delta n$.  
Hence, we find \cite{fnteconv}
\BE
\label{de}
            \delta {\cal E} = \half {\cal C} (\delta P)^2.
\EE

Substituting $\delta P = \psi$ and $\delta {\cal E} = \half {\cal C} \psi^2$ 
into Eqs (\ref{em1}, \ref{em2}) we obtain the nonlinear equations of 
``vacuum hydrodynamics:'' 
\BE
\label{vh1}
\frac{\partial}{\partial t}\left( \chi u_0^2 \right) + 
\frac{\partial}{\partial x^j}\left( \chi u^j u_0 \right) - 
\frac{\partial \psi}{\partial t} = 0,
\EE
\BE
\label{vh2}
\frac{\partial}{\partial t}\left( \chi u_0 u^i \right) + 
\frac{\partial}{\partial x^j}\left( \chi u^j u^i \right) + 
\frac{\partial \psi}{\partial x^i} = 0,
\EE
where $\chi \equiv \psi + \half {\cal C} \psi^2$.   (Note that 
${\cal E} + P = \delta{\cal E} + \delta P = \chi$.)

     These equations represent conservation of energy and momentum, 
respectively, in a fixed reference frame.  Alternatively, one may obtain 
manifestly covariant equations by taking components in the local co-moving 
frame \cite{degroot}; i.e., by considering $u_\mu \partial_\nu T^{\mu \nu}$ 
and $\Delta_{\mu \sigma} \partial_\nu T^{\mu \nu}$, where 
$\Delta_{\mu \sigma} = g_{\mu \sigma} - u_\mu u_\sigma$.  These yield
\BE
\label{ff}
\chi \partial_\mu u^\mu + u^\nu \partial_\nu(\half {\cal C} \psi^2) = 0,
\EE
\BE
\chi u^\nu \partial_\nu u_\sigma - \Delta_{\mu \sigma} \partial^\mu \psi 
= 0.
\EE
Using the first equation, the second can be recast as 
\BE
\partial^\mu(\Delta_{\sigma \mu} \chi) -  
\partial_\sigma (\half {\cal C} \psi^2) = 0.
\EE
Note that, unlike ordinary hydrodynamics, if there is no pressure 
disturbance ($\psi=0$ everywhere), the flow $u^\mu$ is unconstrained.   

     If the pressure disturbance is very small, it is natural to assume, 
provisionally, that the $\psi^2$ terms may be neglected (which has 
the same effect as setting ${\cal C}=0$).  With this simplification 
Eq. (\ref{ff}) reduces to 
\BE
\partial_\mu u^\mu =0
\EE
(wherever $\psi \neq 0$), which is the relativistic version of the 
incompressible-fluid condition.  If one further specializes to 
quasi-1-dimensional solutions (independent of the $y,z$ coordinates) and 
substitutes $u_0=\gamma$, $u=\gamma v$, with $\gamma = 1/\sqrt{1-v^2}$, 
the above equation reduces to 
\BE
\label{relinc}
v \frac{\partial v}{\partial t} +  \frac{\partial v}{\partial x} = 0.
\EE
The other equation for the pressure reduces to
\BE
\label{pres}
 \frac{\partial (\psi v)}{\partial t} +  \frac{\partial \psi}{\partial x} = 0.
\EE
Formally, an implicit solution to (\ref{relinc}) is $v=g(x-t/v)$, where $g$ 
is an arbitrary function (with $g(x)$ representing the initial conditions 
at $t=0$).  The second equation has a formal solution 
$\psi = f(v) \partial v/\partial t$, where $f$ is an arbitrary function.  
These solutions imply a tendency for disturbances to propagate at 
superluminal speeds $1/v$ (see later).  A detailed investigation of the 
solutions to these equations, utilizing some powerful mathematical theory 
\cite{evans}, is a task postponed to future work \cite{fnte2}. 
 
     These strange equations have arisen simply by taking seriously the idea 
that the vacuum is a sort of medium.  The underlying dynamics is unimportant 
except in three respects: (i) it must result in a non-trivial vacuum; 
(ii) it determines the compressibility, ${\cal C}$; (iii) it sets the 
$\mfp$ scale.  In the Standard Model the vacuum is non-trivial both because 
of the quark and gluon condensates of the QCD sector and because of the 
Higgs field.  For the latter case one can reasonably neglect the gauge and 
fermion fields and concentrate on the dynamics of the scalar field alone.  
An account of this dynamics in simple, physical terms has been given by 
Consoli and the author in Ref. \cite{mech} and I outline the story here 
because it makes the notion of the vacuum as a medium much more vivid. 

     Consider single-component $\lambda \phi^4$ theory 
\cite{fnteO4} in a region of parameters where the effective potential has 
both a minimum at $\phi=0$ and a deeper minimum at $\phi=\pm v$.  (Such a 
situation is possible, as will be seen.)  The particle excitations 
above the meta-stable, `empty' vacuum I shall call `phions.'  The 
broken-symmetry vacuum, described in terms of phions, is a spontaneous 
Bose-Einstein condensate, and the Higgs boson is the {\it quasiparticle} 
excitation of this condensate.

    The reason that phions want to condense can be understood in simple terms.
Their fundamental interaction is the 4-point vertex, which, expressed as an 
interparticle potential, is a repulsive $\dtr$ potential.  However, a 
long-range, attractive interaction is induced by the ``fish'' diagram 
involving exchange of two virtual phions.  This interaction has the form 
$-1/r^3$, if one neglects the mass of the exchanged phions.  (Including 
the phion mass basically cuts off this potential at distances 
$\gtrsim 1/m$.)  Contributions from other diagrams either produce short-range 
interactions, which can be absorbed into the $\dtr$ term, or additional 
$-1/r^3$ contributions.
  
     Now consider a large box, volume $\vol$, with periodic boundary 
conditions, that contains $\num$ phions.  Provided the system is dilute 
the ground state corresponds to almost all the phions being Bose-condensed 
in the zero-momentum mode; hence the energy is
\BE
\label{energy}
       E= \num m + \half \num^2 \bar{u},
\EE
where the first term counts the rest energies of the $\num$ phions and the 
second is the number of pairs times the average energy of a pair, $\bar{u}$.  
The diluteness assumption means that three-body interactions, etc., can be 
neglected.  Since almost all the phions are in the zero-momentum mode, whose 
wavefunction is uniform across the box, the average energy of a pair is just
\BE
\label{ubar}
      \bar{u} = \frac{1}{\vol} \int \! d^3r \, U(r),
\EE
where $U(r)$ is the interparticle potential.  Substituting this into 
(\ref{energy}) and dividing by volume gives the energy density as 
\BE
{\cal E} = n m + \frac{n^2}{2} \!\int \! d^3r \, U(r).  
\EE
The potential $U(r)$ contains a repulsive core piece, which integrates to 
a constant (proportional to the scattering length, $a$), and a $-1/r^3$ term, 
which yields $\int \! dr\!/r$.  The ultraviolet divergence can be regulated 
by invoking a core size $r_0$ (inverse cutoff), but an effective infrared 
cutoff $r_{\rm max}$ must also be present.  

    As noted earlier, the phion mass will ultimately cut off the $-1/r^3$ 
interaction at distances $\gtrsim 1/m$.  However, when $m$ is very small a 
more important consideration is the `screening' by the background density of 
phions.  Hence, the $\ln(r_{\rm max})$ basically turns into a logarithm 
of $n$ \cite{fntesc}.  The energy density thus has the form: 
\BE
{\cal E} = \mbox{{\rm sum of}} \,\, n, \,\, n^2, \,\, n^2 \ln n  \,\,\, 
\mbox{{\rm terms.}}
\EE
The three terms represent (i) one-particle rest energies, (ii) the energy cost 
of repulsive, short-range, two-particle interactions, and (iii) the energy 
{\it gain} due to attractive, long-range, two-particle interactions --- with 
the $\ln n$ arising because the incipient infrared divergence is cut off by 
the background density effect.   If the phion mass is sufficiently small then 
(iii) wins, so that an empty box ($n=0$) is energetically disfavoured, 
compared to a condensate of some optimal density $n_v$ (see Fig. 1).  
In field-theory language ${\cal E}(n)$ is $V_{\rm eff}(\phi)$, with 
$n=\half m \phi^2$ \cite{mech}.  

     Next, one wishes to remove the ultraviolet cutoff (send $r_0 \to 0$) 
such that the Higgs mass $M_h$ stays finite \cite{fntefd}.  It turns out 
that the scattering length $a$ associated with the repulsive-core interaction 
(which enters in the coefficient of the $n^2$ term; Cf. Eq. (\ref{dbg})) 
must tend to zero.  In fact \cite{mech}, we need $a$, $m$, and $1/n_v$ to 
be of order $\epsilon^{1/2}$, where $\epsilon = 1/\ln({\rm cutoff}/M_h)$, 
and $M_h^2 = 8 \pi n_v a$.  The vanishing of the scattering length reflects 
the `triviality' of $\lambda \phi^4$ theory in $3+1$ or more dimensions 
\cite{triv}.  Note that {\it `triviality' means hierarchy generation}, since 
it means that the scattering length is tiny (infinitesimal in the limit) 
compared to the physical length scale set by $M_h^{-1}$.  In fact, there 
is a rich hierarchy of length scales \cite{mech}: 
$r_0 \lll a \ll n_v^{-1/3} \ll M_h^{-1} \ll \lambda_{\rm mfp}$.  

    The important point here is that the mean-free-path scale 
$\lambda_{\rm mfp} \equiv 1/(na^2)$ would be infinite if we were to take 
$r_0$ all the way to zero.  Thus, in the infinite-cutoff theory there would 
be no hydrodynamic regime for the vacuum.  However, if the Standard Model 
is only an effective field theory then we can expect $\lambda_{\rm mfp}$ 
to be a long, but finite, length scale.  One natural speculation 
\cite{mech,cpt01} is to identify $a$ with the Planck length --- in which 
case $\lambda_{\rm mfp}$ is of order millimeters/centimeters.  

     I now return to the intriguing feature that the vacuum-hydrodynamic 
equations seem to imply propagation at arbitrarily large speeds $c^2/v$.  The 
first point to make is that hydrodynamics makes sense only for long length 
scales and so is not applicable to sharp wavefronts.  Thus, superluminal 
{\it signalling}, which requires sending real ``news'' (some sort of 
discontinuity), is presumably still impossible.  One only gets in real 
trouble with causality if the propagation speed for arbitrarily high 
frequencies is greater than $c$ \cite{brill}.  The second point is that 
the propagation is of a collective excitation that is made up of phions that 
themselves move at speeds $< c$.  Exactly the same issue arises for a 
non-relativistic Bose-Einstein condensate of atoms:  A phonon with tiny 
momentum $k$ is made up of atoms with tiny velocities $\pm k/m$, yet it 
moves at a finite speed $v_s$.  

    Many other questions remain to be addressed \cite{qhyd}.  To what extent 
are these nonlinear vacuum disturbances akin to sound waves?  How can they 
be detected and/or produced?  Could gravitational-wave detectors be 
sensitive to these scalar, longitudinal disturbances?  What are the 
astrophysical and cosmological consequences?  A thorough study of the 
solutions to the vacuum-hydrodynamic equations may perhaps shed light on 
these and other questions.  

\newpage

\begin{center}
{\bf Acknowledgements}
\end{center}

I am grateful to Maurizio Consoli for many discussions on these topics.  
I also thank Philippe Choquard for illuminating insights into Eqs. 
(\ref{relinc},\ref{pres}).
This work was supported in part by the Department of Energy under Grant No. 
DE-FG05-97ER41031.


\begin{thebibliography} {99}
\bibitem{fnteconv}
     Convexification of the effective potential by the usual Maxwell-type 
construction implies that ${\cal E}(n)$ is constant for $n \le n_v$, so that 
$\delta {\cal E}$ and $\delta P$ vanish when $\delta n$ is negative.  
Eq. (\ref{de}) remains valid but one has the condition $\delta P \ge 0$.
\bibitem{fnte2}
     Remarkably, upon exchanging $x$ and $t$ and replacing $\psi$ with mass 
density $\rho$, these equations become those for an ideal inviscid, 
pressureless fluid; namely the Euler equation, neglecting the pressure 
term (i.e., the inviscid Burgers' equation), and the continuity equation.  
Important progress on this classic problem has been made only recently.  
Choquard \cite{choquard} has pointed out that the condition 
$\rho = \partial v/\partial x$ is necessary to obtain physically 
well-behaved solutions.  This origin of this condition is the determinant 
pre-factor in the semiclassical form of the Feynman propagator.  
The $x \leftrightarrow t$ transformation suggests that our equations will 
be similarly conditioned by the requirement $\psi = f \partial v/\partial t$ 
with constant $f$.
\bibitem{fnteO4} 
    The Standard Model involves a 4-component scalar field, but ultimately 
the three transverse Goldstone fields are ``eaten'' by the vector bosons.  
It is the radial field that is relevant here, so for simplicity I consider 
a single-component theory.  Note that the quanta have no quantum numbers 
and are their own antiparticles.
\bibitem{fntesc}
The propagators of the two exchanged phions suffer multiple collisions with 
background phions.  These collisions act as mass insertions in the 
propagators and modify them to $1/(p^2-M^2)$, where $M^2 = m^2 + 8 \pi n a$.  
The $m^2$ term proves to be negligible, so $r_{\rm max} \sim 1/M 
\propto 1/\sqrt{n}$.
\bibitem{fntefd}
    I alert the reader that this issue is a contentious one.  See the 
introduction of Ref. \cite{mech} for discussion and references to the 
literature.
\end{thebibliography}
\end{document}